\documentclass{ws-mpla_1}

\newcommand{\beq}{\begin{equation}} 
\newcommand{\eeq}{\end{equation}}

\newcommand{\D}{\partial}     
 
\begin{document}     
  
\markboth{J. Bartels and M. Lublinsky}  
{$\gamma^*\,\gamma^*$ scattering via secondary reggeon in QCD}  
  
\catchline{}{}{}{}{}  
%  
%%%%%%%%%%%%%%%%%%%%%%%%%%%%%%%%%%%%%%%%%%%%%%%%%%%%%%%%%%%%%%%%%%%%  
  
\title{$\gamma^*\,\gamma^*$ SCATTERING VIA SECONDARY REGGEON EXCHANGE IN QCD}

\author{\footnotesize JOCHEN BARTELS} 
  
\address{II. Institut f\"{u}r Theoretische Physik, Universit\"{a}t 
Hamburg\\  
Luruper Chaussee 149, 22761 Hamburg, Germany \\  
bartels@mail.desy.de}  
  
\author{MICHAEL LUBLINSKY}  
  
\address{DESY Theory Group, DESY\\  
Notkestr. 85, 22607 Hamburg, Germany \\ 
lublinm@mail.desy.de 
}  
  
\maketitle  
  
%\pub{Received (Day Month Year)}{Revised (Day Month Year)}  
  
\vspace{-10cm} 
\hspace{10cm}  DESY 04-106
\vspace{9cm} 

\begin{abstract}  
We summarize the results on 
the high energy behavior of quark-antiquark exchange in     
$\gamma^*\,\gamma^*$ elastic scattering.  
The ladder diagrams, summed in the double logarithmic approximation, 
provide a perturbative QCD model for secondary reggeon exchange. 
\keywords{QCD; secondary reggeon; quark ladder.}  
\end{abstract}  
  
%\ccode{PACS Nos.: include PACS Nos.}  

\section{Introduction}     
     
$\gamma^*\,\gamma^*$ collisions at high energies provide a unique     
laboratory for testing asymptotic properties of perturbative QCD.     
The virtuality of the photons justifies the use of perturbative      
QCD, and modern electron positron colliders (LEPII, a future linear collider   
NLC) allow to measure the total cross section of $\gamma^*\,\gamma^*$  
scattering  at energies where asymptotic predictions of perturbative QCD  
can be expected to set in.  
The dominant contribution to the process is given by the      
BFKL Pomeron \cite{BFKL} which gives rise to a cross section strongly      
rising with energy $\sigma_{total}^{\gamma^*\,\gamma^*}\sim s^{\alpha_P(0)}$. 
Here $\alpha_P(0)$ is the Pomeron intercept which, in leading order and      
for realistic values of the photon virtualities, lies in the region      
$\alpha_P(0)\simeq 0.3\,-\,0.5$.     
     
There is little doubt that the pomeron will dominate at very high energies,   
and it is expected to be a main contribution at any future linear collider.     
At present, however, the only source     
for experimental data on photon photon collisions is LEP \cite{LEP1,LEP2}.     
These data are at energies which cannot be     
considered as asymptotically large, and it has become clear that at LEP      
energies, the cross section is not yet dominated by the pomeron      
~\cite{BHS,BRL,DDR,MK,NSZ,BFKLP}.      
The data rather indicate the necessity to include, in the theoretical      
description, several corrections. Perturbative corrections      
are due to the quark box (the first $\alpha_s$ correction was computed in 
Ref. \cite{DelDuca}). Nonperturbative contributions include, in particular,  
the soft Pomeron exchange.   
 
Another class of corrections are due to the exchange of secondary reggeons:  
$f_0$ (flavor singlet) or $A_0$, $A_2$ (flavor nonsinglet) ~\cite{DDR,MK,NSZ}. 
In hadron scattering, secondary reggeons denote the exchange      
of mesons and are of nonperturbative nature. 
In virtual $\gamma^*\gamma^*$ scattering, however, we may expect that such  
secondary exchange may become accessible to a perturbative analysis: 
similar to BFKL exchange. The hard scale at both ends of the  
exchanged reggeon provides a justification for using pQCD, provided  
the photon virtualities are sufficiently large. 
If so, meson exchange will be modeled  by the exchange of  
$q \bar{q}$ ladders~\cite{RS}, and the prediction for the energy dependence    may be tested in the corrections to BFKL exchange in $\gamma^*\gamma^*$      
scattering.     
 
Technically speaking, there is a striking difference between gluon exchange  
in the BFKL calculations and quark-antiquark exchange: the appearance of  
double logarithms~\cite{GGL,KiLi}. As result of this, the intercept of the      
$q\bar{q}$-system is of the order      
$\omega_0^{q\bar{q}}\,=\,\sqrt{const\;\alpha_s}$ (as opposed to      
$\omega_0^{BFKL}\,=\,const\;\alpha_s$     
in the single logarithmic high energy behavior of the BFKL Pomeron),     
and its numerical value can be expected to be large. In fact,         
for $q \bar{q}$ scattering it is known ~\cite{RS} that      
the cross section goes as $\sim s^{\omega_0\,-\,1}$ with      
$\omega_0\,=\,\sqrt{2\,\alpha_s\,C_F/\pi}\simeq 0.5$.      
It is remarkable that this intercept obtained in pQCD is very     
close to the nonperturbative one known from Regge phenomenology. 
This observation justifies the hope that a perturbative  
calculation of $q\bar{q}$ exchange in $\gamma^*\gamma^*$ scattering, 
in fact, might be a reasonable model for the exchange of secondary reggeons.  
In this model, pQCD then allows to make an absolute prediction of the  
magnitude of these corrections.     
   
Connected with the appearance of double logarithms in $q \bar{q}$ exchange 
is the role of the infrared region. The ladder graphs that have to be summed  
are infrared finite. However, the comparison with the BFKL approximation  
leads us to expect that there are contribution from small momenta  
which, although not giving rise to infrared divergencies, are not believable.  
In BFKL, this infrared region  
is reached by diffusion in $\ln k_t^2$, i.e. even at large photon  
virtualities $Q^2$ where pQCD is well-justified, the BFKL prediction  
becomes unreliable when $\ln s/Q^2$ becomes of the order $\ln^2 Q^2/\mu_0^2$ 
(where $\mu_0^2$ denotes the infrared momentum scale where pQCD  
becomes unreliable). The double logs in $q \bar{q}$ exchange do not allow for  
diffusion; the region of validity of the pQCD analysis, therefore, is  
different and needs to be investigated. Existing studies of  
fermion-antifermion exchange have never touched 
this question. $\gamma^*\gamma^*$ scattering appears to be a natural  
candidate for addressing this question. 
 
In Ref. \cite{BL} we have performed a detailed study of the high energy
behavior of  
quark-antiquark exchange in $\gamma^*\gamma^*$ scattering in the double  
logarithmic approximation. Three different approaches were exploited in Ref.
 \cite{BL}. Below we will outline only one of them based on a
 Bethe-Salpeter  equation for a sum of Feynman diagrams. 
The other two methods are based on a Mellin space representation of 
the  scattering amplitude. The first one uses the infrared evolution 
equation (IREE) for a partial wave \cite{KiLi}. A third way of handling 
the quark-antiquark exchange utilizes a notion of the reggeon Green's function,
which has been described in Refs. ~\cite{KK,Kirschner}. In Ref. \cite{BL} 
all three methods were shown to agree with each other.

In this letter we present a brief summary of our  
results. We skip most of the technical details of our calculations,  
and we will only outline the way the results are obtained.  
Our main emphasis will be on the physical results.

\section{Integral Equation}     
    
\begin{figure}[th]    
\begin{tabular}{c c c }    
\psfig{file=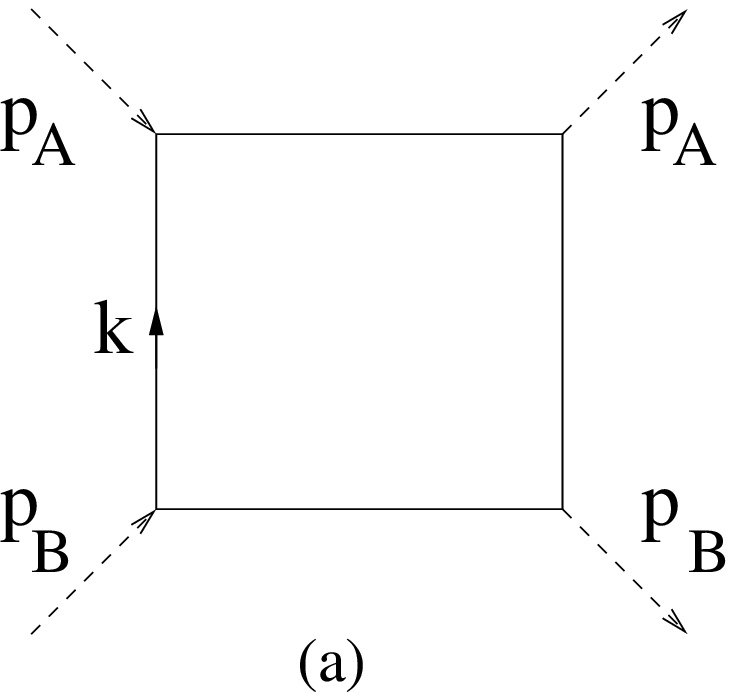,width=40mm}&    
\psfig{file=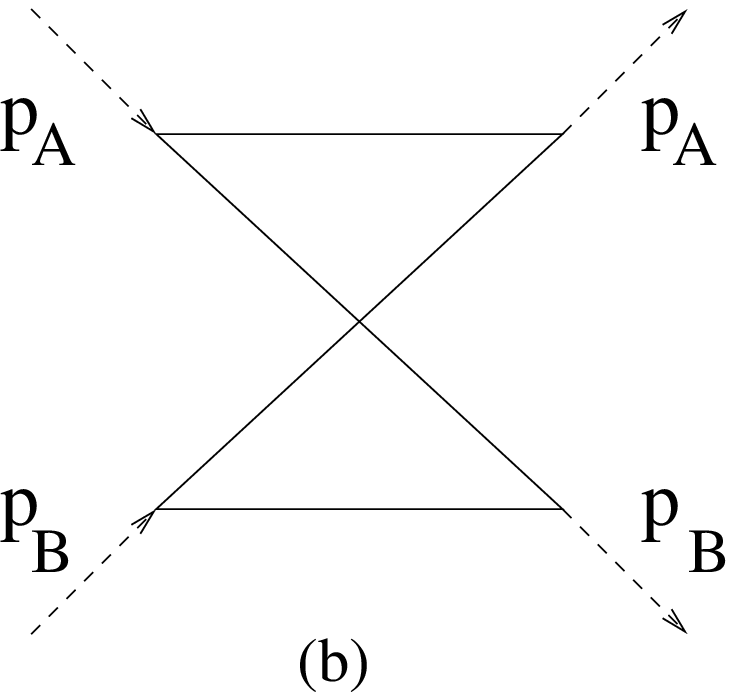,width=40mm}&    
\psfig{file=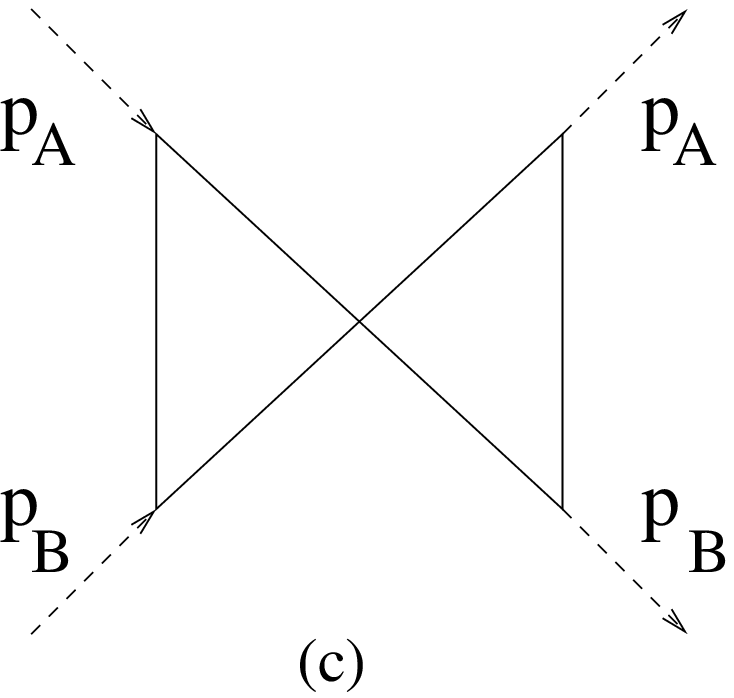,width=40mm} \\    
\end{tabular}   
 \vspace*{8pt} 
  \caption{The Born level diagrams.}    
    \label{born}    
   \end{figure}

We start from the lowest order diagrams for      
the scattering amplitude $T^{\gamma^*\gamma^*}$ of the elastic      
$\gamma^* \gamma^*$ scattering process.  
We consider forward direction $t=0$ only, and for simplicity we first take the  
virtualities  of all external photons to be equal.        
The exact computation of the quark box can be found in Ref.\cite{Budnev};  
we restrict ourselves  to the high energy behavior.      
The lowest order consists of the three fermion-loop     
diagrams (Fig. \ref{born}, a - c);     
however, at high energies the diagram (c) does not contribute to   
the leading double-logarithmic behavior. 
     
The calculations below are done using the Feynman gauge.     
Our method of extracting the double-logs is close to the original      
paper ~\cite{GGL}. With the notation (Fig. \ref{born})      
\begin{equation}     
p_A^2\,=\,-\,Q^2\,;\,\,\,\,\,\,\,\,p_B^2\,=\,-\,Q^2\,;\,\,\,\,\,\,\,\,\,\,     
(p_A\,+\,p_B)^2\,=\,s\,;\,\,\,\,\,\,\,\,\,\,x\,=\,Q^2/s\,.\,\,\,\,\,\,\,\,\,. 
\end{equation}     
and the Sudakov decomposition $k=\beta q - \alpha p + k_{\perp}$      
(with $k_{\perp}^2 = - \vec{k}^2$) the light cone vectors      
$p$, $q$ are defined through     
\begin{equation}     
p_A\,=\,p\,-\,x\,q\,;\,\,\,\,\,\,\,\,\,\,\,\,\,\,     
p_B\,=\,q\,-\,x\,p\,;\,\,\,\,\,\,\,\,\,\,\,\,\,     
p^2\,=\,q^2\,=\,0\,;\,\,\,\,\,\,\,\,\,\,\,\,\,     
2\,(p\,q)\,=\,s\,.     
\end{equation}     
The scattering of the longitudinally polarized photons has additional  
energy suppression, and we will consider transverse polarizations only. 
Transverse polarization vectors are defined by     
$\epsilon_{\pm}^{\mu}\,=\,     
\frac{1}{\sqrt{2}}\, \left(0,1,\pm i,0\right)$.        
      
In order to find a double-logarithmic contribution  
of the quark loop we need to find, in the      
trace expression in the numerator, terms proportional the leading power of      
$s$ and to $k_{\perp}^2$. For the numerator of the planar box diagram  
we obtain:         
\beq    
\label{tr3}     
trace \,\approx \,2 \; s \; k_{\perp}^2 \,\tau_{TT} \,;\,\,\,\,\,\,\,\,\,\,\, 
\,\,\,    
\tau_{TT}\,=\,4\,\,N_c\,\,\alpha_{em}^2\,\, F_{ns} \;\;      
\epsilon(A)\,\cdot \,\epsilon(A') \;\;      
\epsilon(B)\,\cdot\,\epsilon(B'),      
\eeq     
where     
\beq       
F_{ns}\,\,= \,\,    
\sum_{quarks} e_q^4\, \,-\, \,\frac{1}{N_f}\, (\sum_{quarks} e_q^2)^2     
\eeq     
denotes the projection on the flavor nonsinglet t-channel (for the flavor     
group $SU(N_f)$), and     
$e_q$ stands for the electric charge of the quark,  
measured in units of $e$. Diagram Fig. \ref{born}b     
will be obtained by simply substituting, at the end of our      
calculation, $s \to u$.      
     
The analysis of the double logarithmic part of the integration  
region can be summarized as follows. The energies of the s-channel quark lines  
have to be positive; equivalently, the momentum fractions of the large  
incoming momenta range between $x$ and $1$. This leads to the limits: 
\beq 
x \,=\, \frac{Q^2}{s}\, <\; \alpha, \beta \; < \,1\,. 
\label{limit1} 
\eeq 
The momenta of the t-channel quark lines have to be space-like, i.e.  
\beq 
s \,\alpha \,\beta \,< \,\vec{k}^2\,. 
\label{limit2} 
\eeq 
Combining these two conditions we are lead to  
\beq 
\frac{Q^4}{s} \,<\, \vec{k}^2\,. 
\label{limit3}  
\eeq 
Furthermore, in order to justify the use of perturbation theory the  
virtualities of the t-channel quark lines have to be larger than the   
infrared cutoff which we denote by $\mu_0^2$: 
\beq 
\mu_0^2 \,<\, \vec{k}^2\,. 
\label{limit4}    
\eeq    
Now two possibilities arise. For sufficiently large photon virtualities, $Q^2$, 
we have $\mu_0^2 < \frac{Q^4}{s}$, i.e.  
in the entire double logarithmic region the momentum scale of the exchanged  
quark lines is above the infrared cutoff. 
If, on the other hand, the energy $s$ grows and the lower cutoff,  
$\frac{Q^4}{s}$, enters the infrared region $\frac{Q^4}{s}<\mu_0^2$,   
we have to impose a further restriction on the transverse momentum  
integration: $\frac{Q^4}{s} <\mu_0^2 < \vec{k}^2$.  
We therefore define, for the external kinematic variables $s$ and $Q^2$,  
two different regions which we denote by `$+$' and `$-$':     
\beq\label{plusext}     
I^+:\;\; \mu_0^2 \,< \,\frac{Q^4}{s}\,;\,\,\,\,\,\,\,\,\,\,\,\,\,\,\,\,\,  
   \,\,\,\,\,\,\,\,\,\,\,\, 
I^-:\;\; \frac{Q^4}{s} \,<\, \mu_0^2.      
\eeq          
The `+' region defines the hard domain where the result is infrared  
insensitive. Below we will find that, in this region, the amplitude in fact  
has no dependence on $\mu_0^2$ at all.  
 
The integrals are performed in the following way. 
The $\alpha$-integral is done by putting 
one of the s-channel quark lines, say the lowest one, on the mass shell.
This integration sets $\alpha\,=\, x$.  
With the condition that  
\beq 
\vec{k}^2 \,< \,s\,\beta\,, 
\label{limit5}  
\eeq 
the other s-channel quark propagator provides the factor $1/\beta$, 
needed to make the $\beta$-integral logarithmic. 
The remaining integrals are:  
\beq\label{dl}     
\int_x^1 \frac{d \beta}{\beta}\, 
\int_{max(\mu_0^2, \,\beta \,Q^2)}^{\beta \,s} 
\, \frac{d k^2}{ k^2},   
\eeq  
The first energy log then comes from the $\beta$ integration; its limits  
follow from  
(\ref{limit1}). As to the remaining integral over the transverse momentum,  
it is the fermionic nature of the exchanged particles which provides, 
in the trace in the numerator, the additional $k^2$ factor and thus renders  
the momentum integral logarithmic.  
The upper limit of integration results from (\ref{limit5}),  
whereas the lower one  
combines the two possibilities (\ref{limit3}) and (\ref{limit4}). 
If we are in the `+' region, we always have  
$\mu_0^2\, <\, \beta \,Q^2$, 
and there is no $\mu_0$-dependence. On the other hand, 
in the `-' region, we have to split the $\beta$ region into two pieces  
and to perform, for each piece, the $k$ integral separately.         
 
The result for the planar box diagram in the two regions has the form      
\beq     
\label{box1}      
T_{box}^{\pm}\,=\, \,\tau_{TT} \,\left\{     
\begin{array}{ll}     
\ln ^2 \frac{s}{Q^2} & \;\;\;\,\,if\;\;\;\,\,\mu_0^2 \,\,<\,\, 
\frac{Q^4}{s}\\     
 \ln^2 \frac{s}{Q^2} - \frac{1}{2} \left( \ln \frac{s}{Q^2}      
- \ln \frac{Q^2}{\mu_0^2}\right)^2  &\,\, \;\;\; if \,\,\;\;\;    
\frac{Q^4}{s}\,\, <\,\, \mu_0^2     
\end{array} \right\}\,.      
\eeq     
In the second case, $I^-$, we see that we are cutting a piece  
of the phase space, i.e the result for $I^-$ is smaller than for $I^+$.  
The nonplanar box in Fig. \ref{born}b     
is obtained by substituting $s \to u$, and in the      
sum of the two diagrams the obtained results stand for the even signature     
$A_2$ exchange.         
The total cross section for $\gamma^*\,\gamma^*$  
(averaged over the incoming     
transverse helicities) follows from     
\beq     
\label{sigma}     
\sigma^{\gamma^*\,\gamma^*}\,=\,\frac{1}{s}\,Im\,T\,     
\simeq\,\frac{1}{s}\,\frac{\pi\,\D T}{\D \ln s}     
\eeq      
(with the last approximate equality being  
valid in the high energy approximation     
only) and has the form      
\beq     
\label{sigbox}     
\sigma^{\gamma^*\,\gamma^*}_{Born}\,     
=\,\tau_{TT} \,\pi\, \left\{      
\begin{array}{ll}     
2\;\ln \frac{s}{Q^2} & \,\,\,\,\,if\;\;\;\,\,\,\,\,\,\,\,\,\,    
\mu_0^2 \,\,<\,\, \frac{Q^4}{s}\\     
\ln \frac{s}{\mu_0^2}  & \,\,\,\,\,    
if \,\,\,\,\,\;\;\;\,\,\,\,\,\frac{Q^4}{s} \,\,<\,\, \mu_0^2     
\end{array}      
\right\}\,.     
\eeq     
 
It is not difficult to generalize this analysis to the case of unequal      
photon masses, $Q_1^2$ and $Q_2^2$. In this case the boundary of the hard 
domain will be determined from the equation $Q_1^2\,Q_2^2\,=\,\mu_o^2\,s$.

We now turn to higher order corrections to the quark loop diagram.     
For the case of quark-quark scattering it has been shown~\cite{KiLi}      
that the even signature amplitude is described by the QCD ladder diagrams. 
For our analysis we make use of the discussion     
given in ~\cite{GGL,KiLi,Ry,BER2,BER1}.  
       
From the trace expression      
in the numerator we obtain, for each $i$-th rung, a factor     
$k_{i}^2 \,\lambda$ (with  
$\lambda\,\,=\,\,\frac{\alpha_s \,C_F}{2\, \pi}$). Thus each rung in the ladder brings in    
  \beq\label{dl1}     
\lambda\,\int \,\frac{d \,\beta_i}{\beta_i}\,  
\int\, \frac{d \,k_{i}^2}{ k_{i}^2} \,.  
\eeq  
Generalizing the above discussion of the double logarithmic phase space,  
one finds the following ordering conditions: 
$$ 
x\,\, <\, \beta_1\, <\, \beta_2\, <\, ...\, <\, \beta_{n-1}\, 
    < \,\beta_n \,< \,1 \,,
$$ 
$$ 
Q^2\,<\,\frac{k^2_{n}}{\beta_n}\,<\, 
\frac{k^2_{n-1}}{\beta_{n-1}}\, 
<\,...\,<\, \frac{k^2_{1}}{\beta_1}\,<\,s \,.
$$ 
In addition we require for each rung to satisfy 
$k_{i}^2\,>\,\mu_0^2$. 
 
\begin{figure}[th] 
 
\centerline{ 
\psfig{file=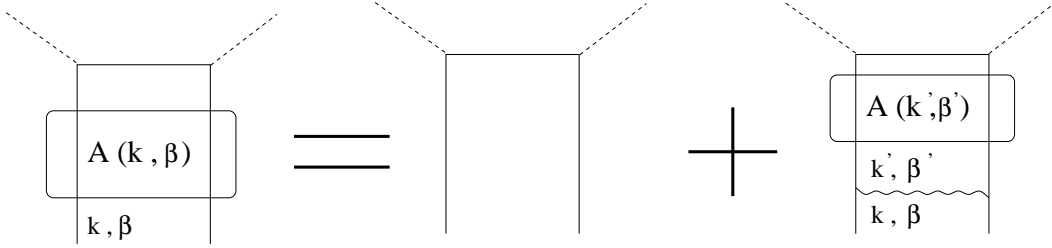,width=140mm} }   
\vspace*{8pt} 
 \caption{ The Bethe - Salpeter equation.}    
    \label{BSfig} 
\end{figure}

Fig. \ref{BSfig} illustrates the Bethe-Salpeter type equation we are  
deriving for the sum of ladder diagrams. We simply remove the lower  
s-channel quark line coupling; in this way we obtain (up to an overall factor)
the     
elastic amplitude $A$ for the photon-quark scattering.  
$A(\beta, \,k^2)\, = \,\sum_n \,A_n$ where  $A_n$ denotes a contribution 
of $n$-rung diagram ($A_0\,=\,1$).  
The amplitudes $A_n$ satisfies the following recurrence relation 
\beq\label{rec} 
A_n(\beta, \,k^2) \,=\, 
\,\lambda\,\int_{Q^2/s}^{\beta} \frac{d \beta'}{\beta'} 
      \int_{max(\beta' \,k^2/\beta\,,\,\,\mu_0^2)}^{\beta'\, s} 
\,\frac{d k^{'2}}{k^{'2}} 
               \, A_{n-1}(\beta', \,k^{'2})\,. 
\eeq 
The limits of integration follow from the above kinematic ordering.       
  
As in the case of the box diagram we have to distinguish between two   
regions $I_A^+$ and $I_A^-$ defined as      
\beq\label{plus}     
I_A^+:\;\;\; \mu_0^2 \,\,< \,\,\frac{Q^2\, k^2}{\beta \,s}     
\,;\,\,\,\,\,\,\,\,\,\,\,\,\,\,\,\,\,\,\,\,\,\,\,   
I_A^-:\;\;\; \frac{Q^2 \,k^2}{\beta \,s}\,\, <\,\, \mu_0^2\,.     
\eeq      
For the amplitudes $A_n$, the kinematics of the lower quark lines plays the  
same role as, in the case of the simple quark box, the lower external photon  
lines. The distinction between the two different regions makes it necessary to  
define two separate amplitudes, $A^+$ and $A^-$. The recurrence relation  
(\ref{rec}) expresses  $A^{-}_n$ in terms of $A^{+}_{n-1}$ and  
$A^{-}_{n-1}$, whereas $A^{+}_n$ only needs $A^{+}_{n-1}$ on the rhs. 
Taking the sum over $n$, we obtain the integral equations     
\begin{equation}\label{BS+}    
A^+(\beta, \,k^2) \,=\,    
1\, + \,\lambda\,\int_{Q^2/s}^{\beta} \frac{d \beta'}{\beta'}    
      \int_{\beta' \,k^2/\beta}^{\beta'\, s} \,\frac{d k^{'2}}{k^{'2}}    
               \, A^+(\beta', \,k^{'2})    
\end{equation}       
and     
\begin{eqnarray}\label{BS-}    
&&A^- (\beta,\, k^2)\,\,=\,\,   1\, +\,\lambda\,    
                \int_{Q^2/s}^{\beta} \frac{d \beta'}{\beta'}    
             \,   \int_{\beta'\, \mu_0^2/Q^2}^{\beta' s}\,   
 \frac{d k^{'2}}{k^{'2}}\,    
                A^+(\beta',\, k^{'2})\,\, \nonumber \\ &  \\    
               &&+\,\,\lambda    
        \int_{Q^2/s}^{\beta \,\mu_0^2/ k^2} \frac{d \beta'}{\beta'}    
                \int_{\mu_0^2}^{\beta'\, \mu_0^2 s/Q^2}   
 \frac{d k^{'2}}{k^{'2}} \,   
                A^- (\beta',\, k^{'2})    \nonumber \\ & \nonumber \\ 
               && + \,\,\lambda\,    
\int_{\beta \,\mu_0^2/k^2}^{\beta}\, \frac{d \beta'}{\beta'}    
                \int_{\beta' \,k^2/\beta}^{s \,\beta' \,\mu_0^2/Q^2}     
                \frac{d k^{'2}}{k^{'2}}\,    
                A^-(\beta', \,k^{'2}) . \nonumber    
\end{eqnarray}    
     
Finally, the amplitudes $T$ for the     
photon-photon scattering is obtained from $A$ by subtracting the      
terms $A_0\,=\,1$, putting $k^2\,=\,Q^2$ and $\beta\,=\,1$ and restoring 
the overall normalization factor.      
Note that in this limit the regions $I_A^+$, $I_A^-$ coincide with      
$I^+$ and $I^-$ in (\ref{plusext}).         
The full scattering amplitude is obtained by adding the twisted     
(with respect to $s \leftrightarrow u$ crossing) fermion loop.       
It is straightforward to generalize  
to the case of unequal photon masses \cite{BL}. 
We only quote the final result for the 
scattering amplitude:   
\beq\label{amp12}     
T^{\pm}(Q_1^2,\,Q_2^2,\,s)\,\,=\,\,    
\left[A^{\pm}(1,Q^2_2)\,-\,1\right]\,\frac{\tau_{TT}}{\lambda}.    
 \eeq    
For the remainder of this letter, we shall discuss the general case of  
unequal photon masses. 
      
\section{Solution of the linear equations}     
     
The structure of the two equations, (\ref{BS+}) and (\ref{BS-})     
defines our strategy: we first solve the equation for $A^+$,  
Eq. (\ref{BS+}),      
and we then use the solution as an inhomogeneous term in the equation for      
$A^-$, Eq. (\ref{BS-}). The main step is a choice of suitable variables.     
         
In the infrared insensitive region $I_A^+$  
where Eq. (\ref{BS+}) holds we define the  
new variables:     
$$\xi\,=\,\ln(\beta\, s/k^2)\,;\,\,\,\,\,\,\,\,\,\,\,\,\,\,\,    
\,\,\,\,\,\,\,\,\,\eta\,=\,\ln     
(\beta\,s/Q_1^2)\,.$$ In these new variables the equation (\ref{BS+})     
can be rewritten      
\beq\label{BS+1}     
A^+(\xi,\,\eta)\,=\,1\,+\,\lambda\,\int^{\eta}_{0}\, d\bar \eta\,    
\int^\xi_{0}\,     
d\bar\xi\,A^+(\bar\xi,\,\bar\eta)\,.     
\eeq      
When differentiated  twice Eq. (\ref{BS+1}) reduces to     
\beq\label{BS+2} \frac{d^2 A^+}{d\xi\,d\eta}\,\,=\,\,\lambda\,A^+ \,.\eeq     
This is a wave equation in the light cone coordinates. Its solution is given by 
\beq\label{sol+1}A^+(\xi,\,\eta)\,=\,I_0 \left     
(\sqrt{4\,\lambda\,\xi\,\eta}\right )\,.      
\eeq     
     
In the infrared sensitive region $I_A^-$   
we have the integral equation (\ref{BS-})      
which couples the functions $A^+$ and $A^-$.     
Let us define a new variable:        
$$\xi^\prime\,=\,\xi\,-\,L_0\,;\,\,\,\,\,\,\,\,\,\,\,\,\,\,\,\,\,\,\,\,\,    
\,\,\,\,\,\,\,\,    
L_0\,=\,\ln (Q^2_1/\mu_0^2)\,.$$     
In the variables ($\xi^\prime ,\,\eta$) the solution of the 
equation (\ref{BS-}) is following     
\beq\label{sol-1}A^-(\xi^\prime,\,\eta)\,=\,I_0\left     
(\sqrt{4\,\lambda\,(\xi^\prime\,+\,L_0)\,\eta}\right )     
\,+\,\frac{\xi^\prime}{\eta+L_0}\,I_2\left     
(\sqrt{4\,\lambda\,\xi^\prime\,(\eta\,+\,L_0)}\right ).     
\eeq

Define     
$$\omega_0\,=\,\sqrt{4\,\lambda}\,;\,\,\,\,\,\,\,\,\,\,\,\,\,\,\,
\,\,\,\,\,\,     
\,\,\,\,\,\,\,\,\tilde Q^2=\sqrt{Q_1^2\,Q_2^2}\,.$$     
     
The amplitude for $\gamma^*\,\gamma^*$ scattering is obtained using     
Eq. (\ref{amp12})      
\beq\label{Tamp+} T^+\,=\,T(Q_1^2,\,Q_2^2,\,s)\,=\,     
\frac{4\;\tau_{TT}}     
{\omega_0^2}\,   
\left[I_0\left(\omega_0\,\sqrt{\ln\frac{s}{     
Q_1^2}\,\ln\frac{s}{Q_2^2}}\,\right)\,-\,1\right]\,     
\,\,\,\,\,if\,\,\, \mu_0^2\,<\,\frac{\tilde Q^4}{s}\eeq     
and     
\begin{eqnarray}\label{Tamp-} T^-&=&T(Q_1^2,\,Q_2^2,\,s)\,=\,    
\frac{4\;\tau_{TT}}{\omega_0^2} \,\times \nonumber \\     
& &\left [     
I_0\left(\omega_0\,\sqrt{\ln\frac{s}{     
Q_1^2}\,\ln\frac{s}{Q_2^2}}\,\right)\,-\;1 - \;\;
\frac{\ln\frac{s\,\mu_0^2}{\tilde     
Q^4}}{\ln\frac{s}{\mu_0^2}}\,    
I_2\left(\omega_0\,\sqrt{\ln\frac{s\,\mu_0^2}{\tilde     
Q^4}\ln     
\frac{s}{\mu_0^2}}\right)\,\right]\,     
\nonumber \\ & &\,\,\,\,\,\,\,\,\,\,\,\,\,\,\,\,\,\,\,    
\,\,\,\,\,\,\,\,\,\,\,\,\,\,\,\,\,\,\,\,\,\,\,\,\,\,\,\,\,\,\,\,\,\,\,\,    
\,\,\,\,\,\,\,\,\,\,\,\,\,\,\,\,\,\,\,\,\,\,\,\,\,\,\,\,\,\,\,\,\,\,\,\,    
\,\,\,\,\,\,\,\,\,\,\,\,\,\,\,\,\,\,\,\,\,\,\,\,\,\,\,\,\,\,\,\,\,\,\,\,    
if\,\,\,\, \mu_0^2\,>\,\tilde     
Q^4/s\,.   
\end{eqnarray}

It is important to note that the     
final result for the amplitude is fully symmetric  
with respect to the photon virtualities.       
The amplitude $T^-$ reduces to $T^+$ when $\tilde Q^4/s = \mu_0^2$, i.e.      
when the dynamical infrared cutoff of the perturbative calculation      
reaches $\mu_0^2$, the limit of the nonperturbative infrared region.     
     
Let us consider, in some more detail, the $s\rightarrow\infty$ asymptotics      
for the case $Q_1^2\,\simeq \,Q_2^2 \gg \mu_0^2$. We take $s$ to be much      
larger than the $Q_i^2$, but still within the region $I^+$ (\ref{plusext}):     
\beq\label{plusregion}     
1 \ll s/\tilde Q^2 \ll \tilde Q^2 /\mu_0^2.     
\eeq     
In this region the asymptotics is obtained from the asymptotic behavior of      
the Bessel function $I_0$      
\beq\label{as+}     
T^+(s\rightarrow     
\infty)\,=\,\frac{4\;\tau_{TT}}{\omega_0^2\,\,\sqrt{2\,\pi\,\omega_0\,\ln     
(s/\tilde Q^2)}}\,\left(\frac{s}{\tilde Q^2}\right )^{\omega_0}\,,     
\eeq     
and the result is entirely perturbative. When $s$ increases and      
eventually reaches the boarder line between $I^+$ and $I^-$:     
$    
s/\tilde Q^2\, = \,\tilde Q^2/ \mu_0^2     
$     
we have to switch to $T^-$. With a further increase of $s$, initially, the      
second term in (\ref{Tamp-}) is not large and we can use the expansion      
of the Bessel function $I_2$ for small arguments. In the asymptotic region     
\beq     
\tilde Q^2/\mu_0^2\, \ll\, s/\tilde Q^2        
\eeq     
the arguments of both Bessel functions are large. The leading asymptotics 
cancel and we have to take into account first corrections. We obtain:     
\begin{eqnarray}\label{as-}      
T^-(s\rightarrow     
\infty)&=&\frac{8\;\tau_{TT}}{\omega_0^2\,\sqrt{2\,\pi}\,(\omega_0\,\ln     
(s/\tilde Q^2))^{3/2}}\,\left(\frac{s}{\tilde Q^2}\right     
)^{\omega_0}\,\times \nonumber \\ & &     
\\ & &\left[1\,+\,\omega_0\,\ln\frac{\tilde     
Q^2}{\mu_0^2}\,+\,\frac{\omega_0^2}{4}\,\ln^2\frac{\tilde     
Q^2}{\mu_0^2}\,+\,     
O\left(      
\frac{\omega_0^4\,\ln^4 \frac{\tilde Q^2}{\mu_0^2} }     
            {\ln \frac{s}{\tilde Q^2}}     
\right)      
\right]\,. \nonumber     
\end{eqnarray}     
     
It is interesting to compare (\ref{as+}) and (\ref{as-}):  
the power behavior      
in $s$ is the same in both regions. The difference lies in the      
preexponential factors: in the second region, we have a slightly stronger      
logarithmic suppression, and there is a logarithmic dependence upon the      
infrared scale $\mu_0^2$.

Another case of interest is deep inelastic scattering on an almost  
real photon     
at very small $x$. This corresponds to the limit      
$ 
\mu_0^2\, \approx Q_2^2\, \ll\, Q_1^2 \,\ll\, s,     
$     
and only the region $I^-$ applies ($Q\equiv Q_1$) with $T_{DIS}(Q^2,\,s)$ 
given by Eq. (\ref{Tamp-}).        
The Bjorken  $x$ is defined in a standard way: $x\,\equiv \,Q^2/s$. The     
flavor nonsinglet     
photon structure function is related to $T_{DIS}$ via     
\beq\label{fns}     
F_{NS}^{\gamma}(x,Q^2)\,     
=\,\frac{Q^2}{4\,\pi^2\,\alpha_{em}}\,\sigma^{\gamma^*\,\gamma}_{tot}\,       
\simeq\,\frac{x}{4\,\pi\,\alpha_{em}}\,     
\frac{\partial \,    
T_{DIS}(Q^2,\,s)}{\partial \ln\,s}\,.     
\eeq     
 
We can consider two different asymptotic limits. The first one is     
$\ln\,1/x\,\gg\,\ln\,Q^2/\mu_0^2\,\gg\,1$.    
 In this limit the structure function     
becomes:     
\beq \label{fns1}     
F_{NS}^{\gamma}(x,Q^2)\,\simeq\,\left( \frac{1}{x} \right)^     
{-1\;+\; \omega_0}\,\,     
\frac{2\;\tau_{TT}\,(\,1\,+\,\omega_0\,\ln (Q^2/\mu_0^2)/4\,)^2}     
{\alpha_{em}\,\omega_0^2\,\sqrt{2\,\pi\,\omega_0}\,     
\ln^{3/2}(1/x) } \,\left(\frac{Q^2}{\mu^2_0}\right )^{\omega_0/2} \,.    
\eeq     
Eq. (\ref{fns1}) gives the Regge limit of the flavor nonsinglet structure     
function.     
Up to the preexponential factor     
this result agrees with the behavior of the flavor nonsinglet     
proton structure function found in ~\cite{Ry,Lu}.     
     
Another asymptotic limit to be considered is     
$1\,\ll\,\ln\,(1/x)\,\ll\,\ln\,(Q^2/\mu_0^2)$ leading to     
\beq \label{fns2}     
F_{NS}^{\gamma}(x,Q^2)\,\simeq\,x\,\,     
\frac{\tau_{TT}}     
{\alpha_{em}\,\pi\,\omega_0^2\,     
\sqrt{2\,\ln (1/x)\,\ln (Q^2/\mu_0^2)}} \,     
e^{\,\omega_0\,\sqrt{\ln (1/x)\,\ln (Q^2/\mu_0^2)}} \,.    
\eeq     
 Eq. (\ref{fns2}) comes from the asymptotic expansion of the first term    
in (\ref{Tamp-}). The second term is subleading in this limit.      
Eq. (\ref{fns2}) corresponds to     
the double logarithmic limit of the DGLAP equation.     
     
\section{The phase space of the double logs}     
     
\begin{figure}[th] 
\begin{tabular}{c c}    
\psfig{file=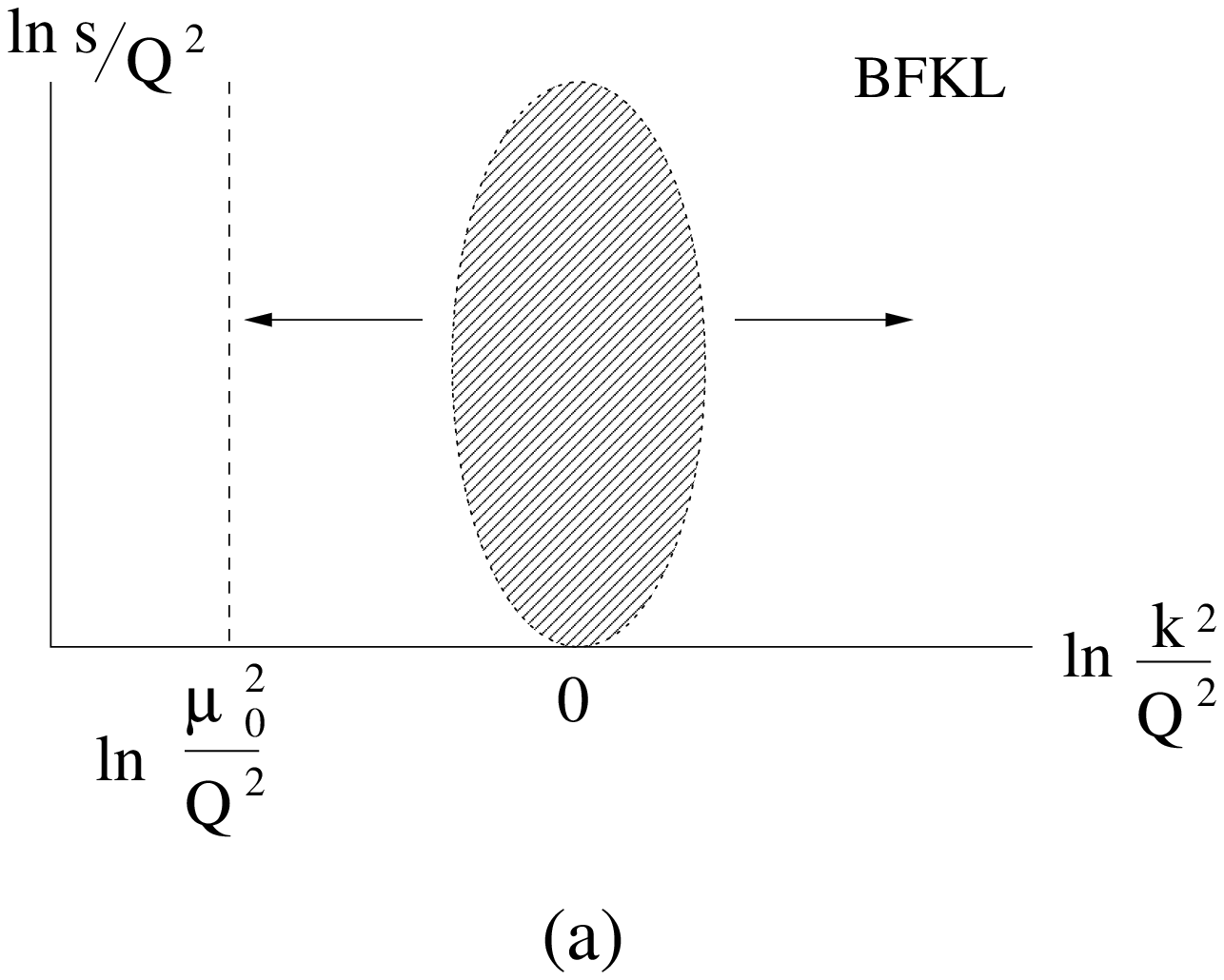,width=60mm}&    
\psfig{file=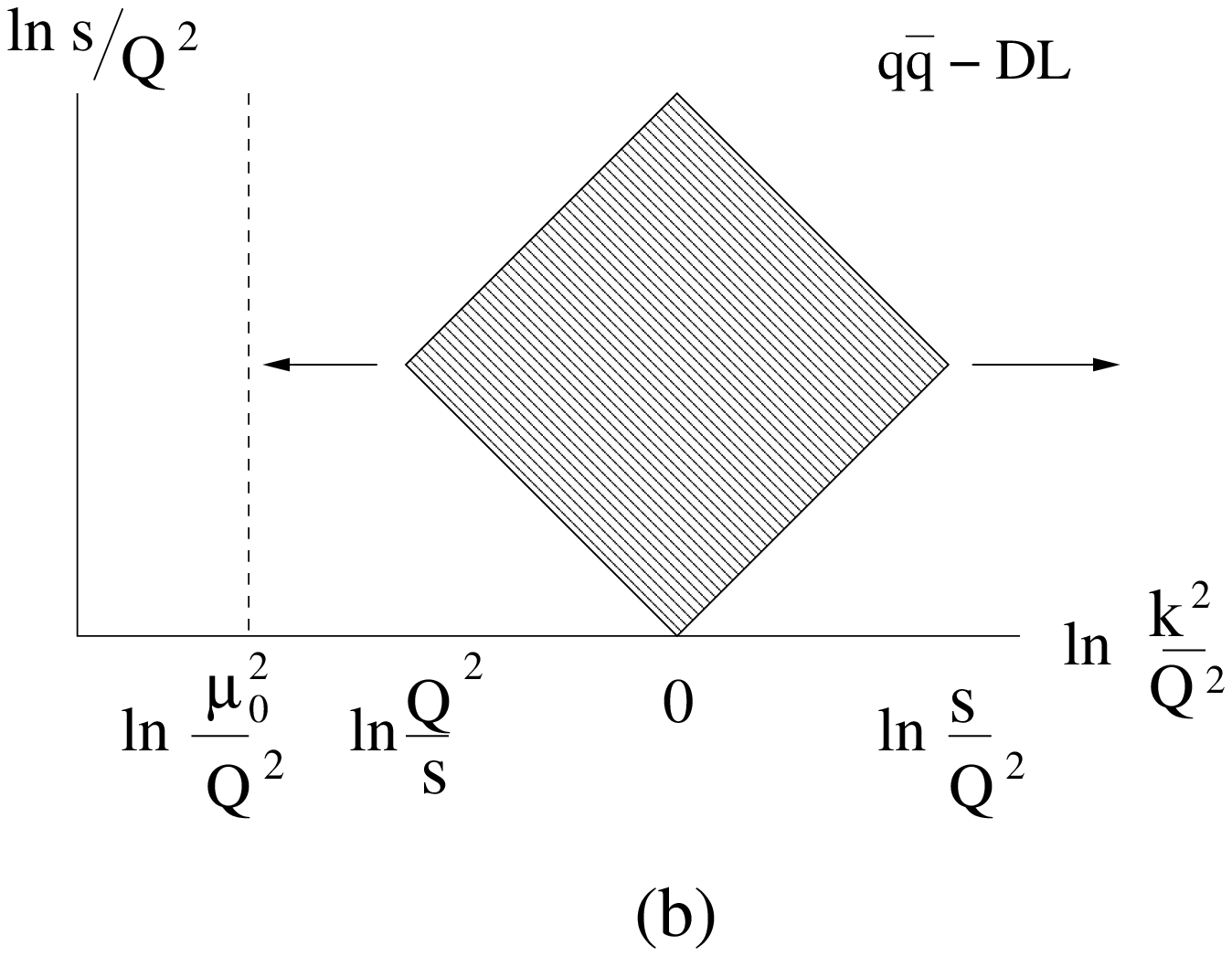,width=60mm}    
\end{tabular}   
 \vspace*{8pt} 
  \caption{Energy dependence of the region    
of integration  in a) BFKL and b) quark ladder.}    
    \label{scalefig} 
  \end{figure}

It is instructive to compare our results for the high energy behavior      
of quark-antiquark exchange in $\gamma^*\, \gamma^*$ scattering       
with those for gluon exchange, i.e. the LO BFKL Pomeron.      
For the latter it is well-known that, for sufficiently large photon      
virtualities and not too high energies, the internal transverse momenta      
are of the order of the photon virtualities and hence justify the use of      
perturbation theory (Fig. \ref{scalefig}a).     
When energy grows, diffusion in $\ln k^2$ (neglecting the running of  
$\alpha_s$) broadens the relevant region of internal transverse momenta,  
which has the shape of a ``cigar''. Its mean size is of the order  
$\sqrt{\ln s}$ and eventually      
reaches the infrared cutoff $\mu_0^2$. If energies increases further,  
the BFKL amplitude -      
although infrared finite - becomes sensitive to infrared physics, and some      
modification has to be included in order take care of nonperturbative  
physics.         
     
With the results of our analysis we now can make an analogous statement     
about quark-antiquark exchange. Since internal transverse momenta range   
between $max(Q^4/s,\,\mu_0^2) \,<\, k^2 \,< \,s$ or, equivalently    
\beq     
max(-\,\ln s/Q^2,\, \ln\,\mu_0^2/Q^2) \,\,<\,\, \ln\, k^2/Q^2\,\, <\,\,  
\ln \,s/Q^2\,,     
\eeq  
there is, for not too large energies, again, a limited region where  
transverse momenta stay above the infrared scale, and the use of     
perturbative QCD is justified. With increasing energy, the $k^2$-region   
expands and eventually hits the infrared cutoff $\mu_0^2$.   
From now on the high energy behavior starts to depend upon infrared physics   
and requires suitable modifications.   
  
In order to understand, in the fermion case, the region of internal   
integration in more detail, let us first return to our amplitude,   
$A^+$ (illustrated in Fig.\ref{BSfig}).  
It satisfies the two dimensional wave equation   
(\ref{BS+2}). It is instructive to introduce the new variables   
\begin{eqnarray}  
\label{time-space}  
t\,=\,\frac{1}{2}\,(\xi \,+\, \eta)\,=\,\ln \frac{\beta s}{\sqrt{k^2 Q^2}}  
\,;\,\,\,\,\,\,\,\,\,\,\,\,\,\,\,\,\,\,\,\,\,\,\,\,\,\,\,\, 
z\,=\,\frac{1}{2}\,(\eta \,-\, \xi)\,=\,\frac{1}{2}\, \ln \frac{k^2}{Q^2},      
\end{eqnarray}  
which leads to the two-dimensional Klein-Gordon equation:  
\beq   
\label{KG}  
\left( \frac{\partial^2}{\partial t^2} \,-\,  
\frac{\partial^2}{\partial z^2} \,-\,   
4 \,\lambda \right)\,\, A^+\; =\; 0   
\eeq  
(here $i \sqrt{4 \,\lambda}$ plays the role of the mass).

Let us confront this with the BFKL Pomeron: in Fig.\ref{scalefig}b   
we have  drawn the square which illustrates the internal region   
of integration. Apart from the difference in shape (``diamond'' versus  
``cigar''),  
the most notable difference is the width in $\ln k^2$. In the fermion case   
it grows proportional to $\ln \,s/Q^2 $, i.e stronger than in the   
BFKL case where $\ln\, k^2$ grows as $\sqrt{\ln \,s/Q^2}$:   
the BFKL diffusion is replaced by a linear growth in the $z$-direction.    
Also the definition of `internal rapidity' is different: in the BFKL case   
the vertical axis can be labeled simply by $\ln \,1/\beta$, whereas in the   
fermion case our variable is  
$t\,=\,\ln \,(\beta \,s / \sqrt{k^2\, Q^2})$ (in both   
cases, the total length grows proportional to $\ln s$).

\section{Numerical estimates}     
     
The final goal of this project should be a confrontation     
of the obtained results  with the LEP data. We are not ready     
yet to produce such a comparison since we still miss     
a significant theoretical contribution arising from     
the flavor nonsinglet exchange. In this section we will present     
some first numerical estimates related to the resummation of the  
quark ladder.     
     
Using Eq. (\ref{sigma})     
the flavor nonsinglet contribution to $\sigma^{\gamma^*\,\gamma^*}_{tot}$     
can be computed from the elastic amplitude (\ref{Tamp+}) and (\ref{Tamp-}).     
In our numerical estimates we will drop the flavor factor $F_{ns}$:     
the missing flavor singlet piece will be estimated to have the same     
functional form as the nonsinglet piece, and (\ref{sigma}) - with the     
factor $F_{ns}$ being replaced by $\sum \,e_q^4$     
- is assumed to represent the sum of     
flavor singlet plus flavor nonsinglet.         
     
\begin{figure}[th] 
\centerline{\psfig{file=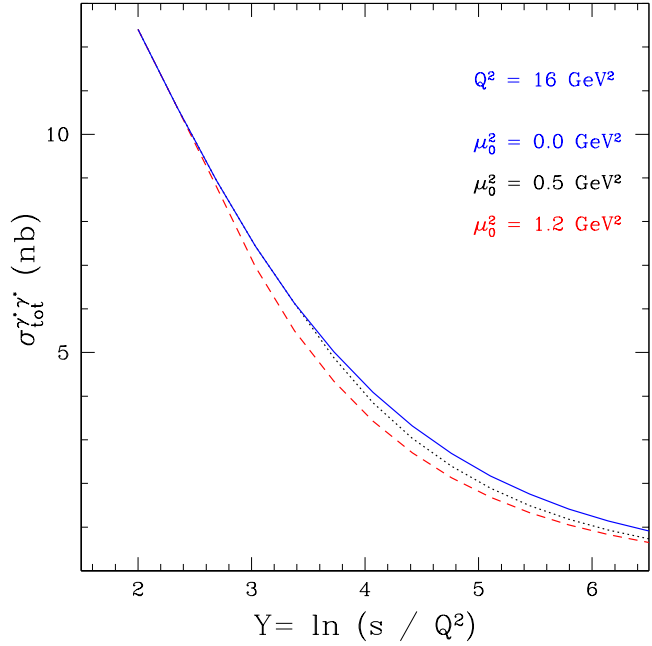,width=90mm}  } 
 \vspace*{8pt} 
 \caption{$q\,\bar q$ contribution to      
$\sigma_{tot}^{\gamma^*\,\gamma^*}$ for various values    
of $\mu_0^2$.}    
    \label{sigfig} 
   \end{figure}

First let us demonstrate numerically the effect of the two kinematical     
regions appearing in  (\ref{Tamp+}) and (\ref{Tamp-}).  
Fig. \ref{sigfig} shows     
the $\gamma^*\,\gamma^*$ cross section as a function of rapidity     
$Y\,=\,\ln s/Q^2$ for equal masses     
$Q_1^2\,=\,Q_2^2\,=\,Q^2\,=\,16\,GeV^2$. Up to $Y\,\simeq 6 $     
this  corresponds to the LEP data region. For the fixed value of $\alpha_s$     
we use $\alpha_s (Q^2)\,\simeq\,0.24$.     
The three curves show the     
dependence of the cross section on the nonperturbative scale $\mu_0$.     
The solid line shows the (unphysical) 
 case $\mu_0^2\,=\,0$ (the region $I^+$),     
the dotted line is $\mu_0^2\,=\,0.5\,GeV^2$, and the dashed line     
is $\mu^2_0\,=\,1.2\,GeV^2$.     
The points where the different curves come together     
correspond to $s\,\mu_0^2/Q^4\,=\,1$. To the left of these points we     
have the hard domain where the perturbative QCD calculation     
is fully reliable and does not depend upon the infrared scale $\mu_0^2$.     
Note that for $\mu_0^2\,=\,0.5\,GeV^2$, almost all     
LEP data are within the hard domain. In this region we should     
expect that the secondary reggeon contribution is described by pQCD, and     
one should not add a further nonperturbative reggeon.     
\begin{figure}[th] 
 
\centerline{      
\psfig{file=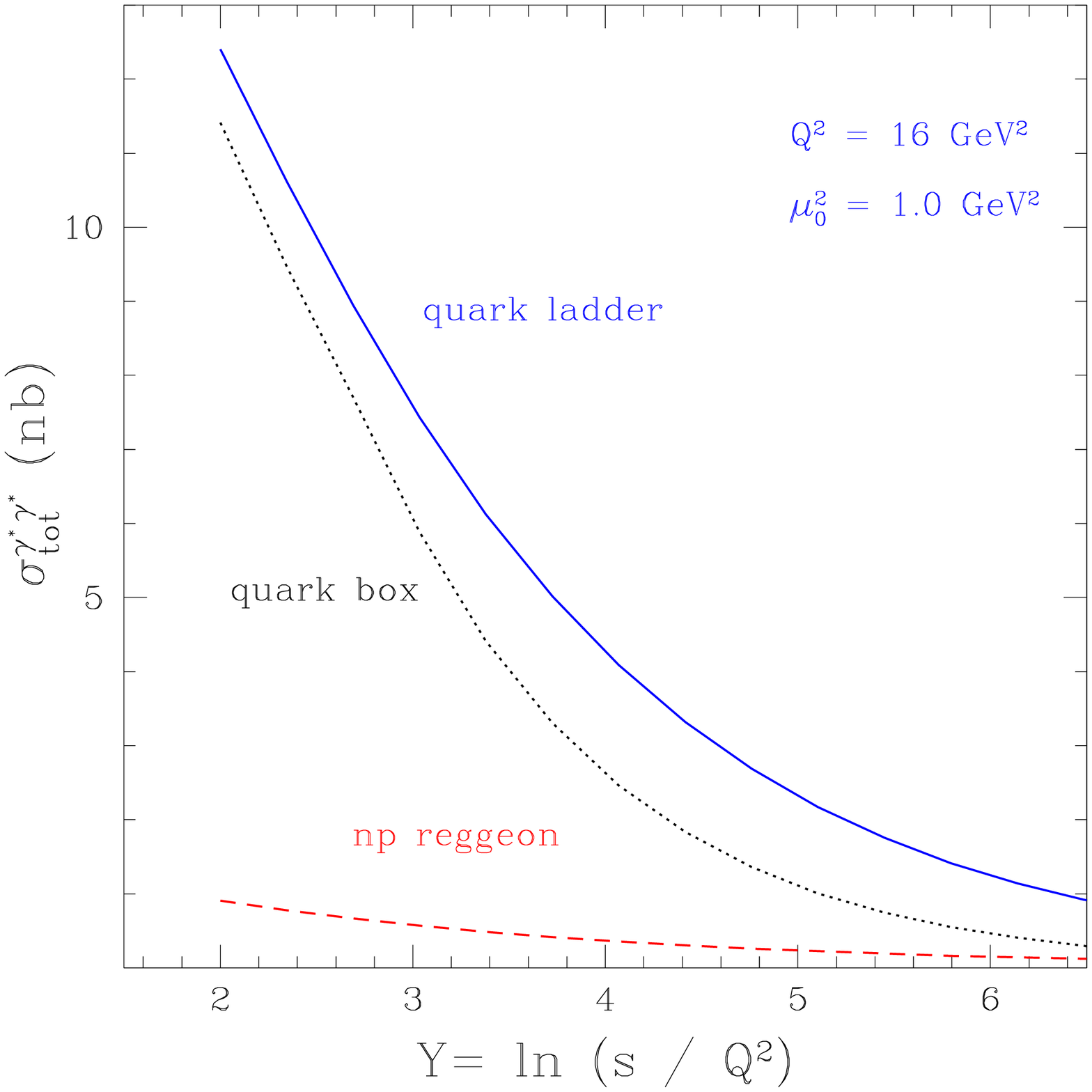,width=90mm}} 
 \vspace*{8pt}   
 \caption{    
Various contributions to $\sigma_{tot}^{\gamma^*\,\gamma^*}$.}    
    \label{bornfig} 
    \end{figure}

Fig. \ref{bornfig} compares the ladder resummation with the box     
diagram contribution\footnote{Only the leading logarithmic contribution is    
taken for the box diagram. The results thus obtained are somewhat larger    
compared to the ones based on the exact expression for the box.}.      
A significant enhancement is observed.     
The enhancement     
grows at higher energies and reaches a factor of ten  at $Y\,\simeq\,10$.     
For comparison we also show the nonperturbative reggeon (dashed line)     
$\sim\,s^{-\,0.45}$.  In Ref. \cite{MK}   
this contribution was added to the box     
diagram in order to fit the data. We believe that within    
the hard domain  our resummed ladder should replace the contribution of the     
phenomenological reggeon. This can be  qualitatively seen from the     
Fig. \ref{bornfig}.    
    
\section{Conclusions}

In this letter we have considered quark-antiquark exchange in     
$\gamma^*\,\gamma^*$ scattering.      
A closed expression for the cross section     
$\sigma^{\gamma^*\,\gamma^*}$ in the flavor nonsinglet channel is given.
The result is valid within the double  logarithmic accuracy of pQCD. 
The cross section depends on the four scales     
relevant for the problem $s\,\gg\,Q_1^2\,\ge\,Q_2^2\,\ge\,\mu_0^2$.     

The resummed quark ladder serves as a model for a `perturbative secondary     
reggeon'. It is     
remarkable that the resulting intercept $\omega_0\,\sim\,0.5 $ is very     
close to the one known from the high energy phenomenology. The large     
intercept is due to the fact that the leading contribution is double     
logarithmic and $\omega_0\,\sim\,\sqrt{\alpha_s}$.

One of the main observations  is the role of the      
infrared cutoff $\mu_0$ introduced into the analysis as the momentum 
scale below which perturbative physics is not reliable.  
For a given energy and for large photon virtualities $Q_1^2$, $Q_2^2$, 
we found that all the internal transverse momenta lie in the hard region. 
For this kinematics the result   does    not 
depend upon the infrared cutoff and the perturbative analysis     
is reliable. When energy increases beyond the value     
$s\,=\,Q_1^2\, Q^2_2 /\mu_0^2$,     
the region of integration penetrates into     
the infrared domain      
and the results starts to depend upon $\mu_0^2$. We show, however,   
that this dependence is logarithmically weak. Another interesting observation
is the role played by $\mu_0^2$  in setting the asymptotic high energy behavior 
of the amplitude. Quite in analogy with the non-forwardness $t$ of the BFKL
physics, the appearance of $\mu_0^2$ modifies the pre-exponential behavior
of the asymptotics.

The study of the quark ladder has an obvious phenomenological motivation.     
The LEP data on $\gamma^*\,\gamma^*$ are at energies at which the quark
 box    still gives a dominant contribution to the cross section.     
We have shown that the gluon radiation leads to a  
significant enhancement of     
the quark box and hence needs to be accounted for.     
The quark box contribution dies fast with energy and is correctly expected     
to be of no importance for $\gamma^*\,\gamma^*$ scattering at a NLC.     
In contrast, the pQCD reggeon receives     
an enhancement of about a factor of ten compared to the quark box,     
and potentially can still give a noticeable correction to the dominant     
pomeron contribution.

We have derived the low-$x$ asymptotics  of the DIS flavor 
nonsinglet photon structure function.  
The power dependence     
on $x$ is shown to be 
the same as for the proton structure function \cite{Ry,Lu}.

In the double logarithmic approximation, the 
intercept of the $q\bar q$ system comes out large. Hence 
it will be important to 
investigate corrections to this leading order result. First corrections 
will come from the single logarithmic contributions of the ladder graphs. 
While the intercept does not acquire a single logarithmic correction 
\cite{KK}
there is a hope that residue corrections  could be computed from the 
reggeon Green's function approach \cite{Kirschner}.  
The influence of the  running strong coupling constant is another 
important aspect expected to come in at the level 
of NLO corrections. Qualitatively, the running coupling enhances the 
importance of the low momentum region. 
 
From the phenomenological point of view our analysis is incomplete.
We have not yet calculated the  flavor singlet     
quark-antiquark exchange which involves an admixture of  
$t$-channel gluons.

\section*{Acknowledgments} 
We wish to thank Boris Ermolaev, Victor     
Fadin, Dima Ivanov, Roland Kirschner, Genya Levin, Lev Lipatov,     
Misha Ryskin, Anna Stasto, and Lech Szymanowski for very     
fruitful discussions.     
     
This research was supported in part by the GIF grant $\#$     
I-620-22.14/1999.


\vspace*{6pt}     
     
     
     
     
\section*{References}    
 
\begin{thebibliography}{0}     
     

\bibitem{BFKL}      
E. A. Kuraev, L. N. Lipatov, and F. S. Fadin, \emph{ Sov. Phys. JETP}     
                {\bf 45}  199 (1977) ; Ya. Ya. Balitsky and L. N. Lipatov,     
                \emph{ Sov. J. Nucl. Phys.} {\bf 28}  22 (1978).     
     
     
     
\bibitem{LEP1}     
P.~Achard {\it et al.}  [L3 Collaboration],    
%``Double-tag events in two-photon collisions at LEP,''    
\emph{Phys.\ Lett.}  {\bf B 531}  39 (2002).    
    
     
\bibitem{LEP2} G.~Abbiendi {\it et al.}  [OPAL Collaboration],        
\emph{Eur.\ Phys.\ J.\ }  {\bf C 24}  17 (2002).    
    
\bibitem{BHS}    
S.~J.~Brodsky, F.~Hautmann and D.~E.~Soper,    
%``Virtual photon scattering at high energies as a probe of the short  distance pomeron,''    
\emph{Phys.\ Rev.\ }  {\bf D 56}  6957 (1997).    
    
\bibitem{BRL}    
J.~Bartels, A.~De Roeck and H.~Lotter,    
%``The gamma* gamma* total cross section and the BFKL pomeron at e+ e- colliders,''    
\emph{Phys.\ Lett.\ }  {\bf B 389} 742  (1996).    
    
\bibitem{DDR}    
A.~Donnachie, H.~G.~Dosch and M.~Rueter,    
%``gamma* gamma* reactions at high energies,''    
\emph{Eur.\ Phys.\ J.\ }  {\bf C 13} 141  (2000);     
%``Two photon reactions at high energies,''    
\emph{Phys.\ Rev.\ }  {\bf D 59} 074011  (1999).     
     
     
\bibitem{MK} J.~Kwiecinski and L.~Motyka,     
\emph{Eur.\ Phys.\ J.\ }  {\bf C 18}  343 (2000).     
     
\bibitem{NSZ}    
N.~N.~Nikolaev, J.~Speth and V.~R.~Zoller,    
%``Color Dipole Bfkl-Regge Factorization And High-Energy Photon Photon Scattering,''    
\emph{J.\ Exp.\ Theor.\ Phys.\ }  {\bf 93}  957 (2001),    
[\emph{Zh.\ Eksp.\ Teor.\ Fiz.\ }  {\bf 93} 1104  (2001)].    
    
\bibitem{BFKLP}    
S.~J.~Brodsky, V.~S.~Fadin, V.~T.~Kim, L.~N.~Lipatov and G.~B.~Pivovarov,    
%``High-energy QCD asymptotics of photon photon collisions,''    
\emph{JETP Lett.\ } {\bf 76} 249  (2002),    
[\emph{Pisma Zh.\ Eksp.\ Teor.\ Fiz.\ }  {\bf 76} 306  (2002)]; 
hep-ph/0111390.       
     
\bibitem{DelDuca} M.~Cacciari, V.~Del Duca, S.~Frixione and Z.~Trocsanyi,    
%``QCD radiative corrections to gamma* gamma* $\to$ hadrons,''    
\emph{JHEP} {\bf 0102}  029 (2001).    
       
   
    
\bibitem{RS} M.G. Ryskin and A.G. Shuvaev,     
\emph{Eur. Phys. J. } {\bf C 25}  245  (2002).     
     
 
\bibitem{GGL} V.~G.~Gorshkov, V.~N.~Gribov, L.~N.~Lipatov and G.~V.~Frolov,     
\emph{Sov.\ J.\ Nucl.\ Phys.\ } {\bf 6} 95 (1968),     
[\emph{Yad.\ Fiz.\ } {\bf 6},     
129 (1967)]; \emph{ Sov.\ J.\ Nucl.\ Phys.\ } {\bf 6}  262 (1968),    
[\emph{Yad.\ Fiz.\ }  {\bf 6}  361 (1967)].    
     
     
\bibitem{KiLi} R.~Kirschner and L.~N.~Lipatov,     
\emph{ Nucl.\ Phys.\ } {\bf B 213}  122  (1983).     
\emph{Sov.\ Phys.\ JETP } {\bf 56} 266  (1982),     
[\emph{Zh.\ Eksp.\ Teor.\ Fiz.\ } {\bf 83}  488 (1982)].     
     
\bibitem{BL} J. Bartels and M. Lublinsky, 
{\it JHEP 0309}  076 (2003).     
     
\bibitem{Ry} B.~I.~Ermolaev, S.~I.~Manaenkov and M.~G.~Ryskin,     
%``Nonsinglet structure functions at small x,''     
\emph{Z.\ Phys.\ } {\bf C 69}  259 (1996).    
     
\bibitem{BER2}     
J.~Bartels, B.~I.~Ermolaev and M.~G.~Ryskin,     
%``Flavor singlet contribution to the structure function g1 at small x,''     
\emph{Z.\ Phys.\ } {\bf C 72}  627 (1996).     
     
     
\bibitem{BER1}     
J.~Bartels, B.~I.~Ermolaev and M.~G.~Ryskin,     
%``Nonsinglet contributions to the structure function g1 at small x,''     
\emph{Z.\ Phys.\ } {\bf C 70} 273  (1996).     
     
     
    
\bibitem{KK}       
J.~ Kwiecinski, \emph{Phys. Rev.} {\bf D 26}   3293 (1982);\\     
R.~Kirschner, \emph{Z.\ Phys.\ } {\bf C 31}  135  (1986).     
     
\bibitem{Kirschner} R.~Kirschner,     
%``Regge asymptotics of scattering with flavor exchange in QCD,''     
\emph{Z.\ Phys.\ } {\bf C 67} 459  (1995).     
     
     
\bibitem{Budnev}     
V.~M.~Budnev, I.~F.~Ginzburg, G.~V.~Meledin and V.~G.~Serbo,     
%``The Two Photon Particle Production Mechanism. Physical Problems. Applications. Equivalent Photon Approximation,''     
\emph{Phys.\ Rept.\ } {\bf 15}   181 (1974).     
       
     
\bibitem{Lu} M. Lublinsky, {\it Phys. Rev.} {\bf D  69}  077502 (2004). 
     
     
    
     
     
\end{thebibliography}
\end{document}